\documentclass[aps,prb,preprint,showkeys,floatfix,byrevtex]{revtex4-1}

\usepackage{bm}
\usepackage{amsmath}
\usepackage{amsfonts}
\usepackage{amssymb}
\usepackage{graphicx}
\usepackage{setspace}
\usepackage{booktabs}

\begin{document}

\title{Re-visiting the O/Cu(111) system -- \\When metastable surface oxides could become an issue!}

\author{Norina A. \surname{Richter}}
\affiliation{Global E3 Institute and Department of Materials Science and Engineering, Yonsei University, Seoul 120-749, Korea}
\author{Chang-Eun \surname{Kim}}
\affiliation{Global E3 Institute and Department of Materials Science and Engineering, Yonsei University, Seoul 120-749, Korea}
\author{Catherine \surname{Stampfl}}
\affiliation{School of Physics, The University of Sydney, NSW 2006, Australia}
\author{Aloysius \surname{Soon}}
\email[Corresponding author. E-mail: ]{aloysius.soon@yonsei.ac.kr}
\affiliation{Global E3 Institute and Department of Materials Science and Engineering, Yonsei University, Seoul 120-749, Korea}

\date{\today}

\begin{abstract}
Surface oxidation processes are crucial for the functionality of Cu-based catalytic systems used for methanol synthesis, partial oxidation of methanol or the water-gas shift reaction. We assess the stability and population of the ``8"-structure, a $|\begin{smallmatrix} 3&2\\ -1&2 \end{smallmatrix}|$ oxide phase, on the Cu(111) surface. This structure has been observed in x-ray photoelectron spectroscopy and low-energy electron diffraction experiments as a Cu(111) surface reconstruction that can be induced by a hyperthermal oxygen molecular beam. Using density-functional theory calculations in combination with \textit{ab initio} atomistic thermodynamics and Boltzmann statistical mechanics, we find that the proposed oxide superstructure is indeed metastable and that the population of the ``8"-structure is competitive with the known ``29" and ``44" oxide film structures on Cu(111). We show that the configuration of O and Cu atoms in the first and second layers of the ``8"-structure closely resembles the arrangement of atoms in the first two layers of Cu$_2$O(110), where the atoms in the ``8"-structure are more constricted.  Cu$_2$O(110) has been suggested in the literature as the most active low index facet for reactions such as water splitting under light illumination. If the ``8"-structure were to form during a catalytic process, it is therefore likely to be one of the reactive phases.  
\end{abstract}

\maketitle

\section{Introduction}
Copper-based systems are used in a wide range of industrial and technological applications, ranging from catalysts used for CO oxidation or the water-gas shift reaction to circuitry wiring to electronic contacts. Many of the various uses of Cu rely on a thorough understanding of the surface properties of the material. Changes in the Cu surface structure due to oxygen adsorption are one important aspect to consider, since surface oxidation is crucial for corrosion processes or for reactions taking place at the surface of a Cu-based catalyst in an oxygen-containing environment. For this reason, great effort has been invested, comprising both, experimental and theoretical approaches, to study the oxidation of the Cu low index surfaces.  Considering the close-packed Cu(111) surface, there is general agreement that oxidation of Cu(111) can lead to substantial reconstruction, up to bulk oxide formation at the surface.\cite{ertl_untersuchung_1967, spitzer_1982, niehus_1983, judd_1986, haase_1988, rajumon_1990, jensen_oxidation_1991, luo_1991, toomes_2000, Xu2001131, Xu2002369, matsumoto_scanning_2001, johnston_structure_2002, soon_oxygen_2006, moritani_reconstruction_2008, zhou_percolating_2008, perez_leon_formation_2012} Clearly, the nature and stability of different partially oxidized surface terminations depend on ambient conditions like partial pressure of the surrounding oxygen gas atmosphere and temperature. In an \textit{ab initio} theoretical study it has been shown that at conditions typical of technical catalysis, the bulk oxide is thermodynamically most stable, followed in the surface energy hierarchy by a thin surface oxide structure.\cite{soon_oxygen_2006} However, it has been pointed out that kinetic effects can play an important role in the surface oxidation process.\cite{soon_oxygen_2006, moritani_reconstruction_2008, zhou_percolating_2008} Depending on temperature and the kinetic energy of oxygen molecules in the vicinity of the surface, different surface reconstructions during the Cu(111) surface oxidation process have been reported. These include the ($\sqrt{13} R46.1^{\circ}\times7R21.8^{\circ}$) structure, known as the ``29" superstructure, and the ($\sqrt{73} R5.8^{\circ}\times\sqrt{21}R10.9^{\circ}$) structure, termed the ``44" structure, observed in combined scanning tunneling microscopy (STM) and low-energy electron diffraction (LEED) experiments at elevated temperatures.\cite{jensen_oxidation_1991, matsumoto_scanning_2001} The designations ``29" and ``44" refer to the large unit cells of these structures, with areas covering 29 and 44 times the $p(1\times1)$ Cu(111) surface unit cell. These structures of the partially oxidized Cu(111) surface were shown to resemble one trilayer of Cu$_2$O(111).\cite{jensen_oxidation_1991, matsumoto_scanning_2001,johnston_structure_2002} Based on an analysis of surface energetics and the electronic and geometric structures of different partially oxidized structures with a $p(4\times4)$ periodicity with respect to the $p(1\times1)$ Cu(111) unit cell, it was shown that the large ``29"- and ``44"-structures can reasonably be modeled with density-functional theory (DFT) methods using the $p(4\times4)$ unit cells, which will be termed $p4$ structures in the following.\cite{soon_oxygen_2006,soon20075809} Further, it was demonstrated that in an oxygen chemical potential region where surface oxides can form, the Gibbs free energies of adsorption of the studied $p4$ oxidized phases lie within a small energy window between $-0.05$ and 0\,eV, indicating that in a situation where thermodynamic equilibrium is not reached and where kinetic effects may play a role, it is likely that different metastable oxidized structures coexist.

\begin{figure*}[tb]
\includegraphics[width=0.85\textwidth]{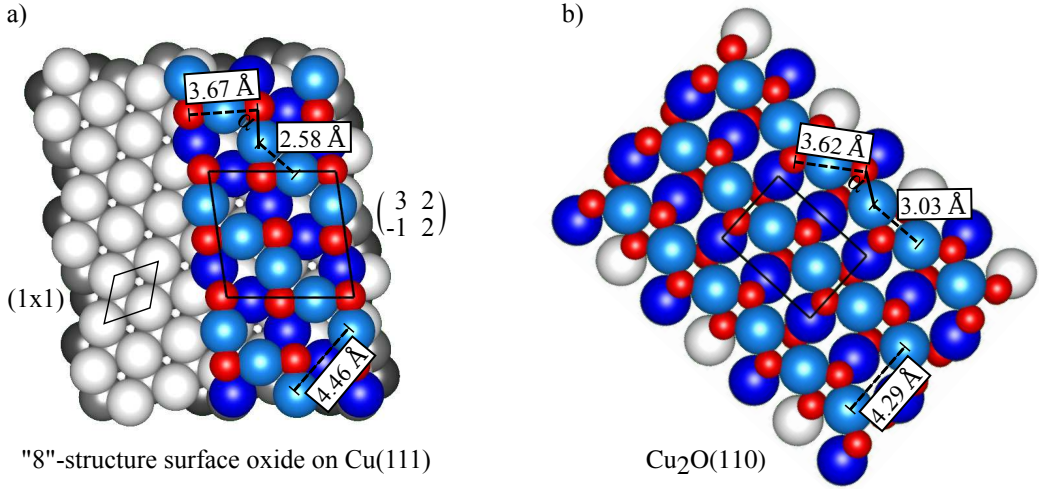}
\caption{(Color online) (a) The ``8"-structure surface oxide on Cu(111) with $|\begin{smallmatrix} 3&2\\ -1&2 \end{smallmatrix}|$ unit cell (right) and the clean Cu(111) surface with $p(1\times1)$ unit cell (left). (b) The Cu$_2$O(110) surface with unit cell. Light blue, dark blue, white, gray and dark gray circles represent Cu atoms in the first, second, third, fourth, and fifth layer, respectively. Oxygen atoms are represented by small red circles. Note that both, the ``8"-structure surface oxide and the Cu$_2$O(110) surface exhibit parallel zigzag-lines of alternating oxygen (red) and Cu atoms in the first layer (light blue). The angle enclosed by one Cu-O-Cu edge is $\alpha=89.1^{\circ}$ for the ``8"-structure on Cu(111) and $\alpha=113.5^{\circ}$ for the Cu$_2$O(110) surface. }
\label{fig1}
\end{figure*}

More recently, another oxidized surface structure on Cu(111) was observed. Using a combination of XPS and LEED measurements at room temperature, Moritani \textit{et al.} showed that exposing the Cu(111) surface to a hyperthermal oxygen molecular beam leads to adsorbate-induced surface reconstruction to a $|\begin{smallmatrix} 3&2\\ -1&2 \end{smallmatrix}|$ structure for an oxygen coverage $\geq\,0.27$ monolayer and translational energies of oxygen $\geq\,0.5$\,eV.\cite{moritani_reconstruction_2008} This structure is not observed, if the incoming oxygen molecules have lower energies. It is argued that for the formation of the $|\begin{smallmatrix} 3&2\\ -1&2 \end{smallmatrix}|$ superstructure, it is crucial that the energy of incoming O$_2$ molecules is large enough to overcome the barrier for dissociative adsorption of O$_2$ at the Cu(111) surface, where the calculated activation barrier for the dissociation is 0.2\,eV.\cite{Xu2001131,Xu2002369} The dissociative processes, in combination with the mechanical energy transfer into the sample due to the collision of atoms at the surface, are suggested to lead to the specific surface reconstruction. The reconstructed surface is metastable; it is reported that annealing to temperatures $\geq\,620$\,K leads to further reconstruction to the well-known ``29"-structure. 

In this work, we analyze the stability and population of the structure observed by Moritani~\textit{et al.}\cite{moritani_reconstruction_2008} using first-principles theory and we discuss this structure with respect to the known low-energy partially oxidized $p4$ structures and in comparison with the Cu$_2$O(110) surface.

\section{Methodology}
The density-functional theory (DFT) calculations in this work are performed using periodic boundary conditions, employing the Projector Augmented Wave method as implemented in the Vienna {\it Ab Initio} Simulation Package (VASP).\cite{kresse_efficient_1996, kresse_ultrasoft_1999} The PBE exchange-correlation functional is used, treating the DFT exchange-correlation energy using the generalized-gradient approximation.\cite{perdew_generalized_1996} All DFT calculations have been tested for convergence with respect to kinetic energy cutoff, {\bf k}-point grid, vacuum separation between repeated slabs, and slab thickness, where total energies and forces do not change more than 20\,meV and 0.02\,eV/{\AA}$^{-1}$, respectively. A planewave kinetic energy cutoff of 500\,eV and a vacuum region of 12\,{\AA} are used for the calculations of the oxidized Cu(111) surface. A Monkhorst-Pack $4\times4\times1$ {\bf k}-point mesh is used for the $p(4\times4)$ and $|\begin{smallmatrix} 3&2\\ -1&2 \end{smallmatrix}|$ unit cells.  The models consist of oxide structures adsorbed on 4 atomic layers of Cu. We relax the coordinates for all structures, keeping the bottom two layers of the Cu substrate fixed, while all other atoms are fully relaxed. Since asymmetrical slabs are used, a dipole correction is employed for all slab calculations. 

Employing \textit{ab initio} atomistic thermodynamics,\cite{weinert_defects_1986, scheffler_1988, reuter_2001, reuter_ab_initio_2005} we calculate the Gibbs free energy of adsorption $\Delta G(p,T)$ according to
\begin{equation}
\Delta G(p,T)=\frac{1}{A}\left(E_{\rm O/Cu}-E_{\rm Cu}-N_{\rm Cu}\mu_{\rm Cu}-N_{\rm O}\mu_{\rm O}\right) \quad,
\label{eq:G_ads}
\end{equation}
where $A$ is the surface area, $E_{\rm O/Cu}$ and $E_{\rm Cu}$ are the DFT total energies of the Cu(111) surface with and without adsorbates, respectively, $N_{\rm Cu}$ is the number of adsorbed Cu atoms, $N_{\rm O}$ is the number of adsorbed O atoms, and $\mu_{\rm Cu}$ and $\mu_{\rm O}$ are the chemical potentials of Cu and O, respectively. The chemical potential of Cu is chosen as the energy of a Cu atom in bulk Cu. The chemical potential of oxygen is calculated with respect to half the energy of an oxygen molecule
\begin{equation}
\mu_{\rm O}(p,T)=\underbrace{\frac{1}{2}E_{\rm O_2}^{\rm tot}\vphantom{\left(\frac{A}{B}\right)}}_{\mu^{\rm ref}_{\rm O}}+\underbrace{\Delta\mu_{\rm O}(p^0,T)+\frac{1}{2}{\rm k}_{\rm B}T\ln\left(\frac{p}{p^0}\right)}_{\Delta\mu_{\rm O}(p,T)} \quad,
\label{eq:mu_O_long}
\end{equation}
where $E_{\rm O_2}^{\rm tot}$ is the total energy of the O$_2$ molecule, $p$ is the partial pressure of oxygen, $p^0$ is the standard pressure $p^0 = 1$\,atm, and $\Delta\mu_{\rm O}(p,T)$ is the chemical potential of oxygen with respect to the reference chemical potential of oxygen $\mu^{\rm ref}_{\rm O}$. To obtain $E_{{\rm O}_2}$, the total energy of a free oxygen atom is calculated using PBE, and the experimental binding energy of an O$_2$ molecule without zero-point energy $E^{\rm exp}_{\rm bind}= 5.22$\,eV\cite{feller_re-examination_1999} is used to correct for a known large error in the PBE binding energy of O$_2$.
$\Delta\mu_{\rm O}(p^0,T)$ can be obtained from tabulated enthalpy and entropy values at standard pressure $p^0 = 1$\,atm. Values used in this work are taken from the JANAF Thermochemical Tables.\cite{janaf} Our analysis does not include the vibrational contribution to the Gibbs free energy of adsorption. The effects of vibrations on the Gibbs free energy of adsorption for surface oxide films on Cu(111) has been estimated based on the Einstein model in earlier work.\cite{soon_oxygen_2006, soon_thermodynamic_2007} According to this estimate, including vibrations does not affect the conclusions drawn in this work.  

The surface population $S(n,p,T)$ for a partially oxidized surface $n$ on Cu(111) within an ensemble of $N$ possible surface structures (at pressures $p$ and temperatures $T$) can be calculated using a Boltzmann distribution, 
\begin{equation}
S(n,p,T)=\frac{e^{\tfrac{-G_n(p,T)}{{\rm k}_{\rm B}T}}}{\sum\limits_{n=1}^{N}e^{\tfrac{-G_n(p,T)}{{\rm k}_{\rm B}T}}} \quad,
\label{eq:surf_population}
\end{equation}
where ${\rm k}_{\rm B}$ is Boltzmann's constant. The numerator in Eq.\,\ref{eq:surf_population} is the probability of observation of structure $n$. The denominator is the canonical partition function. For each structure, the Gibbs free energy of adsorption $G_n(p,T)$ is obtained according to Eq.\,\ref{eq:G_ads}, where we express $\Delta\mu_{\rm O}(p^0,T)$ (cf. Eq.\,\ref{eq:mu_O_long}) as a linear function of temperature in the range between 0 and 1000\,K, extrapolating between the values tabulated in Ref.\,\onlinecite{janaf}.
To obtain a meaningful estimate for the surface population $S(n,p,T)$, the statistical ensemble $N$ has to be large enough and contain the most stable up to high energy surface structures. 

\section{Results and Discussion}
We employ thermodynamics in combination with DFT to calculate the Gibbs free energy of adsorption (cf. Eq.\,\ref{eq:G_ads}) as a measure for the stability of an oxidised Cu(111) surface structure. This assumes that surface and surrounding gas phase atoms are in thermodynamic equilibrium. By construction, kinetic effects are not covered in this approach, however, equilibrium surface phase diagrams can to some extend provide information also on structures that are kinetically stabilized. It has been shown that based on the surface free energies obtained within the equilibrium theory, extended (or so-called metastable) phase diagrams can be constructed.\cite{todorova_handling_2002} For example, it has been demonstrated that for hydrogen adsorption on the Zn-terminated polar ZnO(0001) surface, it is reasonable to consider an extended chemical potential range, going beyond the thermodynamic limit to model the specific experimental situation.\cite{valtiner_hydrogen_2010} Furthermore, if reliable information on the kinetic hinderance of specific low energy surface structures is available, it is possible to use this knowledge to construct a metastable surface phase diagram from the equilibrium phase diagram. By excluding the kinetically hindered surface terminations, information on the next lowest energy surface structures (i.e. the metastable structures) can be revealed.\cite{todorova_handling_2002} The structure we investigate is a $|\begin{smallmatrix} 3&2\\ -1&2 \end{smallmatrix}|$ oxide phase on the Cu(111) surface that was suggested in the literature based on XPS and LEED experiments.\cite{moritani_reconstruction_2008} The relaxed surface as obtained from our calculations is shown in Fig.\,\ref{fig1}a. We term this structure the ``8"-structure, since the area of the surface unit cell is 45.97\,{\AA}$^2$ and therefore almost 8 times as large as that of the underlying Cu(111) surface of 5.85\,{\AA}$^2$. 

It is interesting that the ``8"-structure on Cu(111) resembles the top double layer of the (110) termination of Cu$_2$O, as can be seen by comparing Figs.\,\ref{fig1}a and \ref{fig1}b. Both, the ``8"-structure on Cu(111) and the Cu$_2$O(110) surface exhibit alternating rows of Cu atoms in the second layer (dark blue circles) and Cu atoms in the top layer (light blue circles). The Cu atoms in the first layer build zigzag structures with adjacent O atoms (small red circles) in both surface structures. The distance between O atoms in these zigzag-rows is almost the same in the ``8"-structure and  at the Cu$_2$O(110) surface, namely 3.67\,{\AA} and 3.62\,{\AA}, respectively. Also, the separation between Cu atom rows in the first layer is similar between the two structures, it measures 4.46\,{\AA} for the ``8"-structure and 4.29\,{\AA} for the (110) surface of Cu$_2$O. However, the Cu-O-Cu angle $\alpha$ is 89.1$^{\circ}$ for the ``8"-structure and therefore smaller than the angle enclosed by one Cu-O-Cu edge at the Cu$_2$O(110) surface, where $\alpha$ is $113.5^{\circ}$. This means that the structural features of the ``8"-structure do resemble closely the bulk surface termination of Cu$_2$O(110), however atoms in the thin surface oxide film on Cu(111) are more ``constricted" than at the Cu$_2$O(110) surface, leading to compressive stress in the ``8"-structure oxide film. The close similarity of the ``8"-structure surface oxide layer to top-most layer of the Cu$_2$O(110) bulk oxide surface termination seems to suggest that this ``8"-structure surface oxide may well behave as a precursor bulk oxide growth template -- very similar to that proposed for the earlier observed ``29"- and ``44"-structures.\cite{soon20075809}

\begin{figure}[tb]
\includegraphics[width=0.60\textwidth]{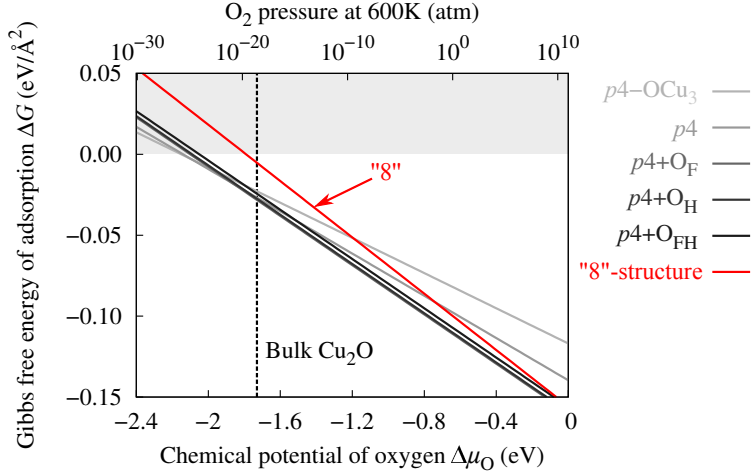}
\caption{(Color online) Gibbs free energy of adsorption for the metastable surface oxide ``8"-structure  on the Cu(111) surface (red line) as a function of oxygen chemical potential, calculated using the PBE exchange-correlation functional. Gray lines show known $p4$ surface structures containing adsorbed oxygen. $\Delta\mu_{\rm O}=0$ corresponds to $\mu_{\rm O}=\frac{1}{2}E(\rm O_2)$, determined from the DFT total energy of the oxygen atom and the O$_2$ binding energy from experiment. The vertical black dotted line indicates the onset of bulk Cu$_2$O formation on Cu(111).}
\label{fig2}
\end{figure}

To clarify if formation of the ``8"-structure is competitive with the earlier observed metastable ``29"- and ``44"-structures, we first calculated the Gibbs free energy of adsorption for the $p4$ structures that were studied in earlier work to model the large surface oxide superstructures.\cite{soon_oxygen_2006,soon20075809} The $p4$ structures are composed of a three-layer lateral unit, where a Cu layer is sandwiched between two layers of O atoms. The basic $p4$ structure is constructed to resemble closely the basic trilayer structure of one layer of Cu$_2$O(111) with a surface unit cell that has hexagonal symmetry to match the available information on the ``29" and the ``44" structures observed in ultra high vacuum scanning tunneling microscopy experiments.\cite{matsumoto_scanning_2001} We use here the same notation as in the original work,\cite{soon_oxygen_2006,soon20075809} where $p4$ denotes the basic hexagonal surface structure, $p4-$OCu$_3$ is the $p4$ structure where one OCu$_3$ unit has been removed, $p4+$O$_{\rm F}$ is the $p4$ structure with an additional O atom adsorbed at the fcc-hollow site, $p4+$O$_{\rm H}$ denotes the $p4$ structure with an additional O atom at the hcp-hollow site, and $p4+$O$_{\rm FH}$ refers to the $p4$ structure with two additional O atoms, one at the fcc-hollow site and one at the hcp-hollow site. Ball-and-stick models of these structures were published in Refs.\,\onlinecite{soon_oxygen_2006} and \onlinecite{soon20075809}. 

Our calculations of the Gibbs free energies of adsorption of the $p4$ structures are in very close agreement, within 0.02\,eV/{\AA}$^2$, with the values reported in Refs.\,\onlinecite{soon_oxygen_2006} and \onlinecite{soon20075809}. In Fig.\,\ref{fig2} we compare the stabilities of these surfaces with that of the ``8"-structure as a function of oxygen chemical potential. In thermodynamic equilibrium, for chemical potentials of oxygen higher than $-1.8$\,eV, bulk Cu$_2$O is stable. For lower oxygen chemical potentials, in the region, where the clean Cu(111) surface is not yet the most stable phase, the adsorption energies of $p4$ surface oxide structures are all within a narrow range and could therefore coexist. However, if formation of the bulk oxide is kinetically hindered, metastable thin surface oxide structures can also form for higher oxygen chemical potentials. We find, that for $\mu_{\rm O}$ close to the oxygen-rich limit, $\Delta\mu_{\rm O}=0$, the ``8"-structure is as stable as the $p4+$O$_{\rm H}$ and the $p4+$O$_{\rm FH}$ structures. In fact, for a fixed oxygen chemical potential $\Delta\mu_{\rm O}$ between -2 and 0\,eV, the maximum difference in Gibbs free energies of adsorption between all considered structures is only 0.05\,eV/{\AA}$^2$. 

Even in the absence of kinetic limitations, the lowest energy partially oxidized surface structure on Cu(111) is not necessarily the (only) one observed in experiment due to a statistic distribution of structures.
In the following we use the calculated stabilities to obtain an estimate of the probability of observing a certain surface morphology and on the resulting population of each surface structure, assuming a Boltzmann distribution of surface structures according to Eq.\,\ref{eq:surf_population}. This concept has recently also been applied to determine the population of metal nanoparticle morphologies.\cite{barnard_clarifying_2014} 
The surface population can be interpreted as the probability of finding a structure out of a statistical ensemble of  partially oxidized surface morphologies on the Cu substrate, assuming that the ensemble includes all possible surface terminations, also comprising structures with high Gibbs free energies of adsorption. 
By construction, in our approach it is not guaranteed that all surface structures are known and accounted for. 

\begin{figure*}[tb]
\includegraphics[width=01.00\textwidth]{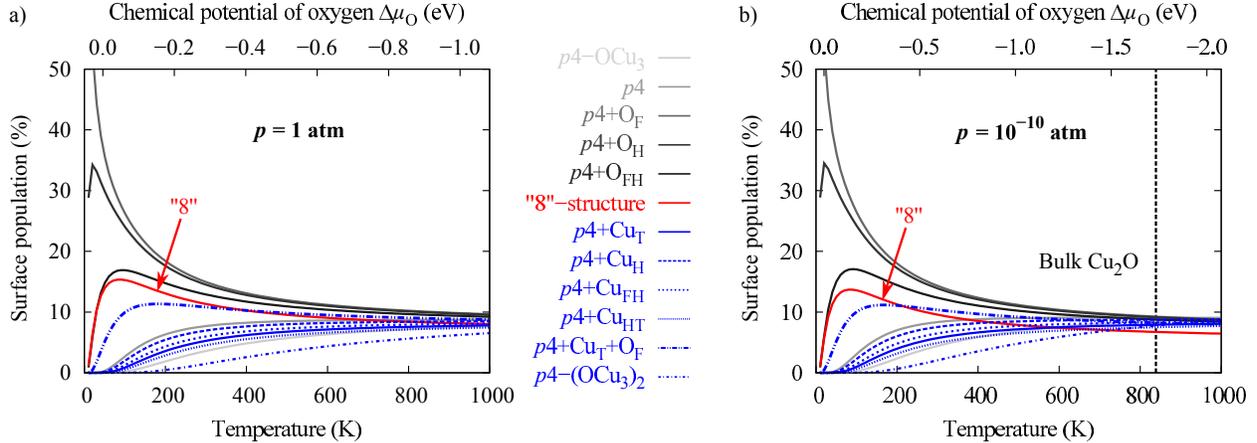}
\caption{(Color online) Surface populations including $p4$-structures and ``8"-structure on Cu(111) as a function of temperature at (a) 1\,atm and (b) 10$^{-10}$\,atm oxygen pressure, assuming a Boltzmann distribution. The vertical black dotted line indicates the onset of bulk Cu$_2$O formation on Cu(111).}
\label{fig3}
\end{figure*}

However, as a reasonable approximation, we include in our statistical ensemble the above mentioned low-energy $p$4 structures, the ``8"-structure, and six more $p$4 structures that have higher adsorption energies: $p4+$Cu$_{\rm T}$, $p4+$Cu$_{\rm H}$, $p4+$Cu$_{\rm FH}$, $p4+$Cu$_{\rm HT}$, $p4+$Cu$_{\rm T}$+O$_{\rm F}$, and $p4-$(OCu$_3$)$_2$, where an index ``T" refers to adsorption on a top site of the Cu(111) substrate. The latter surfaces have been calculated in an earlier work,\cite{soon_oxygen_2006,soon20075809} and we use the corresponding Gibbs free energies of adsorption (according to Eq.\,\ref{eq:G_ads}) to obtain the relative probabilities and surface populations for all structures. We add a correction 0.02\,eV to $\Delta G(p,T)$ for the structures taken from Ref.\,\onlinecite{soon_oxygen_2006}, which corresponds to the difference in $\Delta G(p,T)$ between our calculations and the results reported in Ref.\,\onlinecite{soon_oxygen_2006} for the structures $p4$, $p4+$O$_{\rm F}$, $p4+$O$_{\rm H}$, $p4+$O$_{\rm FH}$, and $p4-$OCu$_3$. The surface populations as a function of temperature for two different oxygen pressures are shown in Fig.~\ref{fig3}. At pressure $p^0 = 1$\,atm, for temperatures $T>500$\,K, which is the region of interest for catalysis applications, the ``8"-structure, the $p4+$O$_{\rm H}$, the $p4+$O$_{\rm FH}$, the $p4+$O$_{\rm FH}$, and the $p4+$Cu$_{\rm T}$+O$_{\rm F}$ structures have statistical populations in the range of 8--15\,\% (Fig.\,\ref{eq:surf_population}b). In fact, all of the considered surface structures on Cu(111) have non-vanishing probabilities of observation, showing that the population of the ``8"-structure is competitive with the known ``29" and ``44" oxide film structures on Cu(111) that are represented by the $p$4 surface phases. However, in the whole discussed temperature and pressure range, the oxygen chemical potential $\Delta\mu_{\rm O}$, shown on the top $x$--axis in Fig.\,\ref{eq:surf_population}, is larger than the enthalpy of formation of Cu$_2$O.

Therefore, we conclude that only if kinetic limitations hinder the formation of the bulk oxide, the ``8"-structure, as well as the $p4$ structures are indeed metastable surface configurations that can coexist according to Boltzmann statistics under relevant catalytic conditions. Obviously, it is possible that some of the surface structures with almost equal populations are preferred or hindered again due to kinetics. However, even only due to the statistic distribution, an oxidized surface phase on Cu(111) that is not the one with highest stability can be observed. At ultrahigh vacuum conditions $p = 10^{-10}$\,atm the surface populations are quantitatively similar to the situation at $p^0 = 1$\,atm (Fig.\,\ref{eq:surf_population}b). However, at temperatures $T > 850$\,K, the bulk oxide is unstable and a coexistence of partially oxidized surface structures on Cu(111) is predicted simply due to the statistic distribution of possible surface terminations.

We have shown that Cu-based catalysts can expose the ``8"-structure as a metastable, partially oxidized phase of the Cu(111) surface, and that this structure resembles the Cu$_2$O(110) surface. Regarding the activity of the ``8"-structure, we note that between the (100), (110), and (111) surfaces of Cu$_2$O, it has been found that the (110) facet is the least stable surface termination and the most photoactive facet for the decomposition of organic materials.\cite{zhang_shape_2010, sun_crystal-facet-dependent_2012, zheng_one-step_2014} This was shown in experiment by studying the crystal-facet-dependent effect of polyhedral Cu$_2$O microcrystals on the photodegradation of methyl orange.\cite{zhang_shape_2010, sun_crystal-facet-dependent_2012} Based on these results it was suggested that Cu$_2$O(110) should be the most active of the three facets for reactions such as water splitting under light illumination.\cite{zheng_one-step_2014} According to our calculations, if the bulk oxide cannot be formed for kinetic reasons, the ``29" and ``44" structures (modeled by the almost equivalent $p4$ structures), which are similar to Cu$_2$O(111), compete with the ``8"-structure, which resembles the Cu$_2$O(110) surface. If these metastable surface oxide films coexist, the ``8"-structure is likely to be among the reactive phases of a Cu-based catalyst for water splitting. The compressed structure of the ``8"-surface, with respect to the similar Cu$_2$O(110) surface, and the resulting lateral stress is a further indication that the ``8"-structure surface oxide film could be quite active towards further reaction in general, including for example, further reconstruction due to oxidation or reactions with different molecules present in the surrounding gas atmosphere. 

\section{Conclusion}
In this work, we have calculated the Gibbs free energy of oxygen adsorption and the surface population for a $|\begin{smallmatrix} 3&2\\ -1&2 \end{smallmatrix}|$ oxidized surface structure on Cu(111). Our results support observations made by XPS and LEED measurements that this ``8"-structure phase can form as a metastable structure, when thermodynamic equilibrium is not reached and kinetic effects hinder the formation of the thermodynamically stable, fully oxidized surface. Due to a statistic distribution of possible surface structures, the ``8"-structure likely coexists with other low-energy phases. We compared the atomic configuration of the ``8"-structure with that of the Cu$_2$O(110) surface and we find that the two systems exhibit very similar structural features, however, in the ``8"-structure the atomic arrangement is more compressed than in the Cu$_2$O(110) surface. The resulting strain could be reduced by further oxidation and reconstruction or by reactions with other species from the environment. This aspect makes the $|\begin{smallmatrix} 3&2\\ -1&2 \end{smallmatrix}|$ structure interesting for surface reactions on Cu-based catalysts.

\section{Acknowledgements}
The authors gratefully acknowledge support by the Basic Science Research Program through the National Research Foundation of Korea (NRF) funded by the Ministry of Education, Science and Technology (NRF Grant No. 2014R1A1A1003415), as well as the Australian Research Council (ARC). Computational resources have been provided by the Korean Institute of Science and Technology Information (KISTI) supercomputing center through the strategic support program for supercomputing application research (KSC-2014-C3-006), and the Australian National Computational Infrastructure (NCI). This work is also supported by the third Stage of Brain Korea 21 Plus Project Division of Creative Materials.

\end{document}